\documentclass{article}
\usepackage{spconf}
\usepackage{amsmath,amssymb,amsfonts}
\usepackage{algorithm}
\usepackage{algorithmic}
\usepackage{graphicx}
\usepackage{textcomp}
\usepackage{booktabs}
\usepackage{multirow}
\usepackage[table,xcdraw]{xcolor}

\begin{document}
\ninept
\title{Study of Adaptive Reweighted Sparse Belief Propagation Decoders for Polar Codes\vspace{-0.5em}}

\name{{Robert M. Oliveira $^{\dag}$ and Rodrigo C. de Lamare $^{\star}$ \vspace{-0.75em}}
\address{ University of York $^{\star}$, UK, \\  PUC-Rio $^{\dag}$, Rio de Janeiro, Brazi\\
Emails:  rbtmota@gmail.com and delamare@puc-rio.br \vspace{-0.995em}}
\thanks{This work was supported by CNPq, CAPES, FAPERJ, and FAPESP. 
}
}

\maketitle

\begin{abstract}
In this paper, we present an adaptive reweighted sparse belief propagation (AR-SBP) decoder for polar codes. The AR-SBP technique is inspired by decoders that employ the sum-product algorithm for low-density parity-check codes. In particular, the AR-SBP decoding strategy introduces reweighting of the exchanged log-likelihood-ratio in order to refine the message passing, improving the performance of the decoder and reducing the number of required iterations. An analysis of the convergence of AR-SBP is carried out along with a study of the complexity of the analyzed decoders. Numerical examples show that the AR-SBP decoder outperforms existing decoding algorithms for a reduced number of iterations, enabling low-latency applications. 
\end{abstract}

\begin{keywords}
 Polar Codes, belief propagation, sum-product algorithm, adaptive reweighting.   
\end{keywords}

\section{Introduction}
Increasing polar codes \cite{Arikan,pdshort,nupol,piecega} decoding speed is an important area of research motivated by the performance requirements of the $5^{th}$ generation wireless networks \cite{5G}. The first decoder proposed for polar codes was the successive cancellation (SC) \cite{Arikan} decoder, which has low complexity and modest performance. However, the SC decoder is characterized by serial decoding, which is an error-prone decoding strategy. Because of this, SC decoders often have low performance for applications that require high-speed real-time decoding with low latency. The successive cancellation list (SCL) decoding \cite{Tal} is an alternative to the SC decoder that yields a significant performance improvement. In its operation, it stores the most likely codewords in a list, reducing the probability of error. Furthermore, by concatenating a cyclic redundancy check code \cite{Tal}, SCL can be further improved. The decoding of the list of successive cancellations aided by the cyclic redundancy check (CRC) has a performance comparable to LDPC and Turbo codes \cite{Tal}. One major disadvantage of SC and SCL decoding is a long decoding latency due to their serial decoding nature. Since these decoding techniques do not directly provide soft-in/soft-out information, they are not suitable for iterative decoding and detection \cite{spa,mfsic,mbdf,bfidd,dfcc,1bitidd,did,dynovs,aaidd,listmtc,detmtc,dynmsgamp,dynmsgamp2,comp,llraps} applications without modifications.

The belief propagation (BP) decoder is an alternative to the problems mentioned above. The BP decoder has high parallelism, high throughput and low latency \cite{Arikan2}. However, for high performance it requires a large number of iterations. List decoding BP \cite{Elkelesh} can be used to enhance performance. It is employed when the standard BP graph does not produce a successful decoding, so the permuted version of the graph can generate improved estimates. To improve the convergence of the BP decoder, a reweighting technique based on the Euclidean distance was presented in \cite{Lucas} along with a Q-Learning algorithm to adjust the reweighting parameter. 

BP decoding is performed over a factor graph that corresponds to the $\textbf{G}_N$ generator matrix, which is dense with many short cycles. Cammerer \cite{Cammerer} introduced an alternative formulation called low-density parity-check code (LDPC)-like polar codes decoder, which is an application of the sum-product algorithm (SPA) for decoding polar codes. 
It employs a pruning technique that transforms the { dense \textbf{G} matrix into a sparse \textbf{H} matrix}, and uses systematic encoding to maximize performance. However, its performance is worse than the standard BP and degrades further for long blocks. Ebada \cite{Ebada} proposed a performance improvement in terms of bit error rate (BER) and frame error rate (FER) with the polar codes construction optimized for LDPC-like decoding by genetic algorithms. The performance obtained is comparable to SCL for short to medium blocks but gets worse for long blocks. Chen \cite{Chen} proposed a polar decoding method based on both the Layered BP and the modified Node-Wise Residual BP (NW-RBP) scheduling strategies. The performance obtained is comparable to that of SCL even for long blocks.

In this paper, we propose an adaptive reweighted sparse belief propagation (AR-SBP) decoder for polar codes. The AR-SBP decoding algorithm is inspired by decoders that employ the sum-product algorithm for LDPC codes. In particular, the AR-SBP decoder introduces reweighting of the exchanged log-likelihood-ratio in order to refine the message passing, improving the performance of the decoder and reducing the number of required iterations. An analysis of the convergence of the AR-SBP algorithm is developed along with a study of the complexity of AR-SBP and existing decoders. Numerical examples show that the proposed AR-SBP decoder outperforms existing decoding algorithms for a reduced number of iterations, enabling low-latency applications. 

This paper is organized as follows. Section 2 reviews the fundamentals of polar codes. Section 3 describes the polar decoder. Section 4 presents the proposed AR-SBP decoder and Section 5 shows an analysis of its convergence. Section 6 presents and discusses the simulation results. In Section 7 we draw the conclusions. \vspace{-1em}

\section{Polar Codes}

Given a symmetric binary-input, discrete and memoryless channel (B-DMC) $W:\mathcal{X} \to \mathcal{Y}$, where $\mathcal{X}$  $\in \{0,1\}$ and $\mathcal{Y} \in \mathbb{R}$. We have that $W(y|x)$ is the channel transition probability, with $x \in \mathcal{X}$ and $y \in \mathcal{Y}$. In order to transmit the information bits, the most reliable sub-channels are chosen. $\mathcal{A}$ is the set of $K$ indices. In turn, $\mathcal{A}^\text{c}$ is the complementary set, containing the indices of the least reliable channels where the frozen bits are placed. Polar codes can be completely specified by three parameters, PC$(N,K,\mathcal{A}^\text{c})$, where $N$ is the block length with code bits, $K$ is the number of information bits and the code rate is $R=K/N$.

Let us denote $W^N: \mathcal{X}$ $^N$ $\to \mathcal{Y}$ $^N$ with
\begin{equation}
  W^N(y_1^N|x_1^N)=\prod_{i=1}^N W(y_i|x_i) 
\label{eq01}
\end{equation}

The mutual information is defined by \cite{Arikan}
\begin{equation}
 I(W) = \sum\limits_{y \in Y}\sum\limits_{x \in X}\frac{1}{2}W(y|x)\log\frac{W(y|x)}{\frac{1}{2}W(y|0)+\frac{1}{2}W(y|1)},
\label{eq02}  
\end{equation}
where the base-2 logarithm $0 \leq I(W) \leq 1$ is employed. 
On the $N$ independent channels of $W$ we apply the polarization process \cite{Arikan}, we obtain a set of polarized channels $ W_N ^ {(i)}: \mathcal{X} \to \mathcal{Y} \times \mathcal{X}^{\text{i-1}}$, $i = 1,2,\ldots,N$. As defined in \cite{Arikan}, this channel transition probability is given by
\begin{equation}
  W_N^{(i)}(y_1^N,u_1^{(i-1)}|u_i) = \sum\limits_{u_{i+1}^N \in X^{N-1}}\frac{1}{2^{N-1}}W_N(y_1^N|u_1^N). 
\label{eq03}   
\end{equation}

According to \cite{Arikan}, $N \to \infty$, {$I(W_N^{(i)})$} tends to $0$ or $1$.

The encoding is given by ${\mathbf x}^N_1 = {\mathbf u}^N_1\textbf{G}_N$, where $\textbf{G}_N$ is the transformation matrix. ${\mathbf u}^N_1 \in \{0.1 \}^N$ is the input block. ${\mathbf x}^N_1 \in \{0.1 \}^N$ is the codeword, where ${\mathbf u}^N_1 = [{\mathbf u}_{\mathcal{A}},{\mathbf u}_{\mathcal{A}^ \text{c}}]$, with ${\mathbf u}_{\mathcal{A}}$ are bits of information and ${\mathbf u}_{\mathcal{A}^\text{c}}$ are frozen bits. We define $\textbf{G}_N = \textbf{B}_N\textbf{F}^{\otimes n}_2 $, where $\otimes$ denotes the Kronecker product, $\textbf{F}_2 = \footnotesize\left[\begin{array}{cc}
1 & 0 \\
1 & 1 \end{array} \right]$, $\textbf{B}_N$ is the bit-reversal permutation matrix and $n =\log_2 N$. A simplification without loss of generalization is the omission of $\textbf{B}_N$.

\section{LDPC-like polar decoder}

A {sparse} matrix $\textbf{H}$ for polar codes can be constructed from the corresponding generating {dense} matrix $\textbf{G}_N$ \cite{Cammerer}. The conventional factor graph of the BP decoder is converted to the bipartite Tanner graph similar to LDPC codes \cite{bfpeg,memd}. Then a pruning process is applied to make the graph sparse \cite{Cammerer}. Sparse means that the number of zeros in the matrix $\textbf{H}$ is much higher than the number of ones. The rows of $\textbf{H}$ are called check nodes (CNs) and the columns of $\textbf{H}$ are called variable nodes (VNs). Such codes are often represented graphically by a Tanner graph \cite{Fossorier}. Fig. 1 shows a Tanner graph and its $\textbf{H}$ matrix.

\begin{figure} [htbp]
\begin{center}
\includegraphics[scale=0.9]{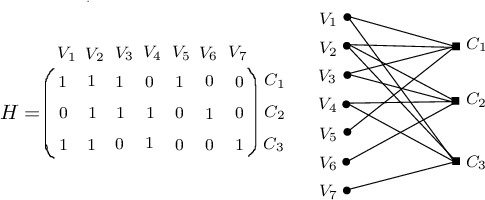}
\caption{Tanner graph and parity check matrix \textbf{H}} 
\vspace{-1 em}
\end{center}
\label{fig01} 
\end{figure}

The LDPC-like polar decoder is a message-passing decoder with iterative processing over the factor graph, in which the LLRs about the coded bits are exchanged along the edges of the Tanner graph. First, each VN gets its corresponding LLR in the received codeword, which we refer to as the VN’s intrinsic LLR. Each VN represents a bit in the codeword. In the beginning, VNs send their LLRs to their connected CN. Each CN processes all the messages sent from its connected VNs, and then computes a value for each VN. Each VN then gets all these computed values from its connected CNs, adds them all together and to its intrinsic LLR, and obtains more precise LLRs. This concludes one iteration. If the stopping criterion is not met, each VN computes and sends its so-far computed LLRs to each CN and the iterations continue until the stopping criterion is met. As the algorithm proceeds, the computed LLRs for each VN gets closer to a {fixed LLR value}. This procedure is continuous and is only interrupted under two conditions: when decoding is successful and the code is found or when $T_{max}$ is reached. Note that with only locally available information it is possible to process information in the CNs and VNs, which allows an efficient and parallelizable decoding.

In terms of notation, if the $\text{c}^{th}$ CN is connected to the $\text{v}^{th}$ VN, the message from that CN to that VN is given by $\Lambda_{c \to v}$, and the message in the opposite direction is $\lambda_{v \to c}$. At the start of the decoding, the messages are initialized as follows: 
\begin{equation}
\Lambda_{c \to v} = \textbf{0},
\label{eq11}
\end{equation}
\begin{equation}
\lambda'_{v \to c} = \textbf{y},
\label{eq12}
\end{equation}
where $\lambda'_{v \to c}$ is the initial condition. Each iteration of the decoder executes the following three steps: CN update, VN update and stop criterion.

For the CN update, all the CNs produce their messages to all their connected VNs. If the number of VNs is V then the message from the $\text{c}^{th}$ CN to the $\text{v}^{th}$ VN can be computed with
\begin{equation}
\Lambda_{c \to v} = 2 \cdot {\rm tanh}^{-1} \left( \prod_{k=1,v \ne v'}^V  {\rm tanh} \left(\frac{\lambda_{v \to c}}{2}\right) \right).
\label{eq13}
\end{equation}
The notation ($v'\neq v$) in $\eqref{eq13}$ denotes that to compute the message to a certain VN, the messages from all other connected VNs is taken into account, except for the message that has come from that VN.

In the VN update, all the VNs produce their messages to all their connected CNs. The number of CNs is C, the message from the $\text{v}^{th}$ VN to the $\text{c}^{th}$ CN is computed according to 
\begin{equation}
\lambda_{v \to c} = \lambda'_{v \to c} + \sum_{k=1}^C \Lambda_{c \to v}.
\label{eq14}
\end{equation}

The stopping criterion \cite{Zhang} at each iteration takes into account that a more precise codeword is expected to be achieved. The estimate $\hat{\textbf{x}}$ at the end of each iteration is determined as
\begin{equation}
\hat{\text{x}}=\begin{cases}  1, \ \lambda_{v \to c} \leq 0 \\ 
 0, \ \lambda_{v \to c} > 0.
 \end{cases}
\label{eq15}
\end{equation}

If the stopping criterion is not met then $\lambda'_{v \to c}$ is updated by
\begin{equation}
\lambda'_{v \to c} = \lambda_{v \to c},
\label{eq16}
\end{equation}
which indicates that we have new CN and VN updates, and a new stopping criterion. 

If the stopping criterion is met then $\lambda'_{v \to c}$ is updated with the new codeword
\begin{equation}
\lambda'_{v \to c} = \textbf{y},
\label{eq17}
\end{equation}
and a new set of steps begins.

\section{Proposed adaptive reweighted sparse BP algorithm}

In this section, we propose an adaptive reweighted technique for message passing decoders that can reduce the number of required iterations. 
The proposed AR-SBP decoder is based on the difference of values of $\lambda_{v \to c}$ between the current and the previous iterations.
{
The reweighting factor, $\rho \in (0,1]$, is then applied in the VN and must ensure the convergence of the AR-SBP algorithm to a fixed point \cite{Roosta}. 
When $\rho = 1$, we have the standard BP message passing algorithm. 
As we can see in eq. $\eqref{eq14}$, the LLR update is applied to the VN nodes. Therefore, the reweighting process consists of assessing how LLRs evolve in terms of signal and modulus over time in the VN messages updates.  

As the node updates depend on the signal and modulus operators, it is undeniable the importance of these operators on the reweighting process as well. Thus, based on eq. $\eqref{eq14}$, we introduce three parameters, namely, the edge weight $\rho$, the general correction factor $\beta$, and the correction factor of the adjustment direction $\Delta$, which will adaptively modify how the LLRs are updated.

Rewriting $\eqref{eq14}$ and including a reweighting factor $\rho$, we obtain
\begin{equation}
\lambda_{v \to c} = \rho* \left(\lambda'_{v \to c} + \sum_{n=1}^N \Lambda_{c \to v}\right),
\label{eq18}
\end{equation}
where $\rho$ is the associated edge weight, and
\begin{equation}
\rho = 1 - \beta \cdot \left[\frac{||\lambda_{v \to c}|-|\sum_{n=1}^N \Lambda_{c \to v}||}{(|\lambda_{v \to c}|+|\sum_{n=1}^N \Lambda_{c \to v}|)}\right]\cdot \Delta,
\label{eq19}
\end{equation}
where $\Delta$ is correction factor of the adjustment direction, or increment or decrement, with
\begin{equation}
\Delta = {\rm sign}\left(\lambda_{v \to c} + \sum_{n=1,v \ne v'}^N \Lambda_{c \to v} \right).
\label{eq20}
\end{equation}
The condition $v'\neq v$ denotes that to compute the message to a certain VN, the message from all the other connected VNs is taken into account, except for the message that has come from that VN. The parameter $\beta$ suggests the general correction factor. We initially consider $\beta$ equal to 1. In the VN updates, the reweighting is based on the distance between the LLRs of $|\lambda_{v \to c}|$ and $|\lambda'_{v \to c}|$, and whether the signals have changed over the iterations. Note that when $|\lambda_{v \to c}|$ and $|\lambda'_{v \to c}|$ are close to each other, $\rho$ is approximately equal to 1. Thus, the VN update is similar to that of  $\eqref{eq13}$. Besides that, the reweighting responds to the value deviation between $|\lambda_{v \to c}|$ and $|\lambda'_{v \to c}|$ and the signal correction by $\Delta$. The parameter $\beta$ is a general factor, which according to our studies should belong to the range $(0, 1]$. Thus, an open problem is how to set up the most appropriate $\beta$ for a specific input. In a similar reweighting scenario for standard BP \cite{Arikan}, the work in \cite{Lucas} adjusts $\beta$ with the help of the Q-learning algorithm.
The proposed AR-SBP decoding algorithm is detailed in Algorithm 1, where $\textbf{y}$ and $\text{T}_{max}$ are inputs and $\hat{\textbf{x}}$ is an output vector.

\begin{algorithm}[h]
 \caption{Adaptive reweighted sparse BP decoder}
 \begin{algorithmic}[1]
 \renewcommand{\algorithmicrequire}{\textbf{Input:}}
 \renewcommand{\algorithmicensure}{\textbf{Output:}}
 \REQUIRE Maximum loop, $\text{T}_{max}$
 \REQUIRE Sparse matrix of $\textbf{G}_N$, $\textbf{H}$
 \REQUIRE Number of CNs, C
 \REQUIRE Number of VNs, V
 \REQUIRE Output channel LLR signal, $\textbf{y}$
 \ENSURE  Estimated binary codeword, $\hat{\textbf{x}}$
 \STATE T = 0
 \FOR {i = 1:CN}
 \FOR {j = 1:VN}
 \STATE $\lambda'(\text{i,j})_{v \to c}=y(\text{j})*H\text{(i,j)}$
 \ENDFOR
 \ENDFOR
 \REPEAT
 \STATE T = T + 1
 \FOR {i = 1:CN}
 \FOR {j = 1:VN}
 \STATE $\Lambda(\text{i,j})_{v \to n} = $
 \STATE $2 \cdot {\rm tanh}^{-1} \left( \prod_{k=1,k \ne \text{j}}^V  {\rm tanh} \left(\frac{\lambda'(\text{i},k)_{v \to c}}{2}\right) \right)$
 \ENDFOR
 \ENDFOR
 \FOR {j = 1:VN}
 \FOR {i  = 1:CN}
 \STATE $\Delta(\text{i,j}) = {\rm sign}\left(\lambda(\text{i,j})_{v \to c} + \sum_{k=1}^C \Lambda(k,\text{j})_{c \to v}\right)$ \\ 
 \STATE $\rho(\text{i,j}) = 1 - \beta  \left[\frac{||\lambda'(\text{i,j})_{v \to c}|-|\sum_{n=1}^N \Lambda(\text{i,j})_{c \to v}||}{(|\lambda'(\text{i,j})_{v \to c}|+|\sum_{n=1}^N \Lambda(\text{i,j})_{c \to v}|)}\right] \Delta(\text{i,j})$ \\ 
 \ENDFOR
 \STATE $\lambda_{v \to c} = \rho* \left(\lambda'_{v \to c} + \sum_{n=1}^N \Lambda_{c \to v}\right)$
 \STATE $\hat{x}(\text{j})=\begin{cases}  1, \ \lambda(\text{i,j})_{v \to c} \leq 0 \\ 
 0, \ \lambda(\text{i,j})_{v \to c} > 0,
 \end{cases}$ 
 \ENDFOR 
 \FOR {i = 1:CN}
 \FOR {j = 1:VN}
 \STATE $\lambda'(\text{i,j})_{v \to c}=\rho(\text{i,j}) \cdot \lambda(\text{i,j})_{v \to c}$
 \ENDFOR
 \ENDFOR 
 \UNTIL $\text{T}=\text{T}_{max} \ \text{or} \ \hat{\textbf{x}}\textbf{H}^T=0$
 \RETURN $\hat{\textbf{x}}$
\end{algorithmic}
\end{algorithm}

\section{Convergence analysis}

The proposed AR-SBP algorithm is an iterative algorithm: the statistical information in the decoder is updated with each message pass. In the graphical model of the decoder, these updates ensure convergence at each node. In the standard BP algorithm, messages are adjusted by weights based on edges determined by the graph structure in which all weights are unitary. The work in \cite{Wainwright} demonstrates that suitable choices of these weights converge to a single point in the graph. Moreover, it has the benefit of dealing with convex problems, resulting in more stable message passing updates \cite{Roosta}. 

According to \cite{Roosta}, the convergence of the reweighted BP algorithm can be guaranteed by any of the following explicit conditions: row sum condition and column sum condition. Thus, the convergence of the reweighted BP algorithm depends on the correct choice of the edge weight $\rho$.

For the row sum condition, we have
\begin{equation}
\max_{(v \to c)} \left( \sum_{u \in V\setminus c} \rho_u + (1-\rho_V) \right) \lambda^n_{v \to c} < 1,
\label{eq21}
\end{equation}
where the term $\rho_V = \sum^V \rho$. The work in \cite{Roosta} considers all $\lambda_{v \to c}$ normalized ($\lambda^n_{v \to c}$), i.e. $\sum^V \lambda^n_{v \to c} = 1$. Since all $\lambda^n_{v \to c} < 1$, we have
\begin{equation}
 1+\sum_{u \in V\setminus c} \rho_u - \sum^V \rho = 1-\rho_c<1,
\nonumber
\end{equation}
and since $\rho_c < 1$, eq. \eqref{eq21} holds.

For the column sum condition, we have
\begin{equation}
\max_{(v \to c)} \left\{\rho_C\left(\sum_{u \in C} \Lambda_{u \to c} \right) + (1-\rho_C) \Lambda_{v \to c} \right\} < 1, 
\label{eq22}
\end{equation}
where $\rho_C$ is the reweighting parameter associated with the check nodes. If we consider all $\rho_C=1$ the work in \cite{Roosta} considers all $\lambda_{v \to c}$ normalized ($\lambda^n_{v \to c}$), i.e., $\sum^V \lambda^n_{v \to c} = 1$. According to eq. \eqref{eq13}, then all $\Lambda_{v \to c}<1$, and eq. \eqref{eq22} is verified.

In turn, the reweighting parameter $\rho$ has an adaptive characteristic depending on the absolute values of the LLRs. We notice that its value tends to $\gamma$ with the evolution of the iterations, since $(||\lambda_{n \to m}|-|\sum_{n=1,n \ne n'}^N \Lambda_{m \to n}||)$ tends to zero, which yields
\begin{equation} 
\begin{split}
\lim_{T\to\infty} \rho &= \lim_{T\to\infty} \gamma - \left[\frac{||\lambda_{v \to c}|-|\sum_{n=1,v \ne v'}^N \Lambda_{c \to nv}||}{(|\lambda_{v \to c}|+|\sum_{n=1,v \ne v'}^N \Lambda_{c \to nv}|)}\right]\cdot \Delta = \gamma
\end{split}
\vspace{-0.25em}
\label{eq23}
\end{equation}
where $\gamma$ is set to 1. The factor $\rho$ is adjusted to accelerate convergence. Note also that $\rho$ does not increase the complexity of the BP decoder, which remains $\mathcal{O}(2\text{T}_{max} \log N)$ \cite{Wang}. \vspace{-0.5em}

\section{Simulations}

In this section, we present simulation results to evaluate the performance of the AR-SBP decoding algorithm introduced in Section 4. The proposed AR-SBP decoder is applied to polar codes to obtain faster convergence, which we illustrate with numerical examples that show the performance versus the average number of iterations or versus the signal-to-noise ratio (SNR) defined as ${\rm SNR} = E_b/N_0$ \cite{Vangala}. In addition, we also show a comparative analysis of the proposed AR-SBP decoder with other competing decoders. In the examples, the channel is additive white Gaussian noise (AWGN) and the design-SNR \cite{Arikan} is set to 1dB.

In Table 1 we show the average number of iterations for the polar code designs PC(128,64), PC(256,128), PC(512,128) considering the BP, NW-RBP and the proposed AR-SBP decoders. We notice that the AR-SBP decoder presents lower average number of iterations than the sparse BP decoder, and close to the NW-RBP decoding algorithm. The AR-SBP decoding algorithm has a convergence close to that of the NW-RBP decoder, whereas the AR-SBP decoder is much more efficient from a computational viewpoint than the NW-RBP decoder. Indeed, NW-RBP requires the implementation of a step of search and classification. A reduction in the number of iterations in the order of $60\%$ is observed for $E_b/N_0 >$ 3dB with the NW-RBP and the proposed AR-SBP decoders. \vspace{-0.85em}
\begin{table}[htb]
\centering
\caption{Average number of iteration, $\text{T}_{max}$=20.} \vspace{-0.15em}
\begin{center}
\begin{tabular}{|c|c|c|c|c|c|}
\hline 
N                    & $E_b/N_0$(dB)    & 1      & 2         & 3         & 4 \\ 
\hline \hline
\multirow{3}{*}{128} & BP               & 20.00  & 20.00     & 18.00     & 15.23 \\ 
                     & AR-SBP           & 19.11  & 12.05     & 8.71      & 7.31 \\ 
                     & NW-RBP           & 19.09  & 10.59     & 8.41      & 7.25 \\ 
\hline                     
\multirow{3}{*}{256} & BP               & 20.00  & 20.00     & 17.13     & 13.46 \\ 
                     & AR-SBP           & 19.32  & 12.15     & 9.13      & 7.33 \\ 
                     & NW-RBP           & 19.21  & 11.05     & 8.75      & 7.29 \\ 
\hline                     
\multirow{3}{*}{512} & BP               & 20.00  & 20.00     & 16.74     & 12.92 \\ 
                     & AR-SBP           & 19.48  & 12.01     & 9.23      & 7.25 \\ 
                     & NW-RBP           & 19.47  & 11.32     & 8.55      & 7.35 \\ 
\hline                     
\end{tabular}
\end{center}
\end{table}

In Fig. 2, we assess the BER performance of the PC(256,128) design with $E_b/N_0=2dB$ for the BP, NW-RBP and AR-SBP decoders. We notice that AR-SBP is superior to BP and very close to NW-RBP. In this scenario, from the sixth iteration on, the AR-SBP and NW-RBP decoders have equivalent performance.

\begin{figure}[htbp]
\vspace{-1 em}
\begin{center}
\includegraphics[scale=0.5]{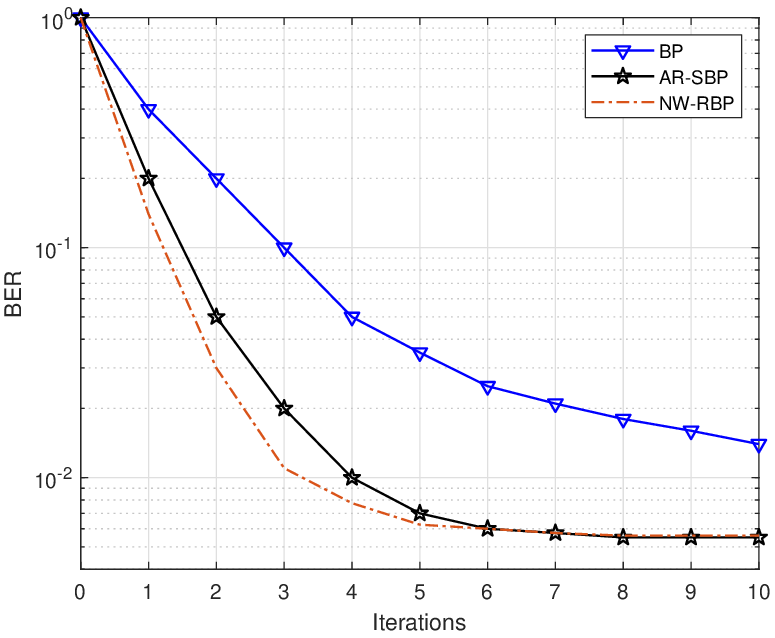}
\vspace{-1.2em}
\caption{BER versus iterations for BP, AR-SBP and NW-RBP.}
\end{center}
\vspace{-1em}
\end{figure}

In the Fig. 3 we have a comparative analysis of the performance of the polar codes $N=256$, rates $R=1/2$ and for the SC, BP, SCL and AR-SBP decoders. For the BP \cite{Arikan2} decoder we use $\text{T}_{max}=60$. For the SCL decoder \cite{Tal} we use the list with size $\mathcal{L}$=128, this value of $\mathcal{L}$ being very close to the maximum likelihood ($\mathcal{L}=256$ for $N=256$). In addition, for the AR-SBP decoder we use $\text{T}_{max}=20$. Note that for $N=256$ we adopted the design of polar codes based on a genetic algorithm \cite{A.Elkelesh}, which generates the same $\mathcal{A}^c$ as the Gaussian approximation construction \cite{Chung}.

 We notice that the performance of the AR-SBP decoder is better than that of the SC and BP decoders, and comparable to that of the SCL decoder. The BP decoder presents low performance due to a large number of short cycles when compared to the AR-SBP decoder. As for the SCL decoder, even considering the size of the list, its performance is similar to AR-SBP. At this rate the set $\mathcal{A}^c$ is large enough to ensure a high rate of successful AR-SBP decoded codewords. As for the computational cost of the decoding algorithms \cite{Wang} we have for BP $\mathcal{O}(2\text{T}_{max}\log N)$, for SCL $\mathcal{O}(\mathcal{L}N\log N)$ and for AR-SBP $\mathcal{O}(2\text{T}_{max}\log N)$, being the AR-SBP the least complex decoding algorithm together with BP. 

\begin{figure}[htb]
\vspace{-1.2em}
\begin{center}
\includegraphics[scale=0.5]{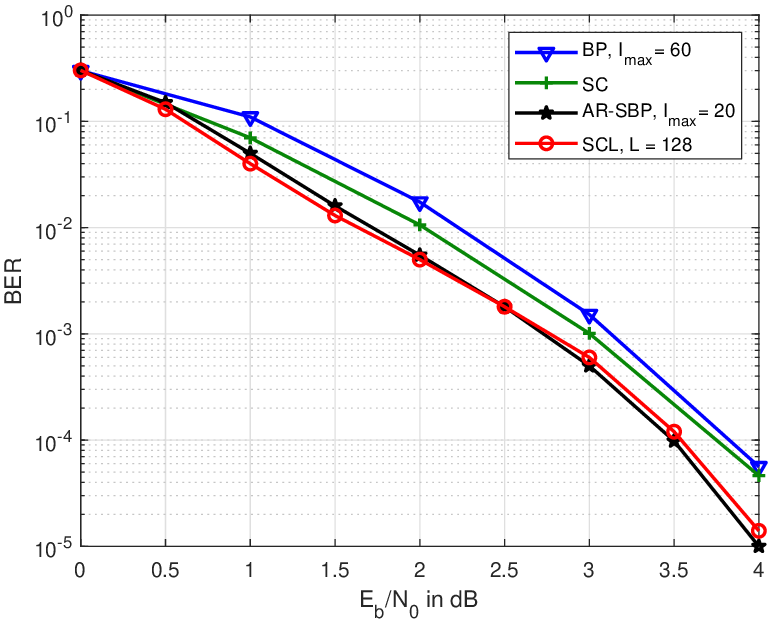}
\vspace{-1.2em}
\caption{Polar codes performance for $N=256$ and $R=1/2$, for BP, AR-SBP and SCL, for $\text{T}_{max}$=60, $\text{T}_{max}$=20 and $\mathcal{L}$=128, respectively.}
\end{center}
\vspace{-1 em}
\end{figure}\vspace{-1em}

\section{Conclusions}

In this work, we have proposed an AR-SBP decoder for polar codes which improves the decoding performance while reducing the number of iterations. 
The proposed AR-SBP algorithm accelerates the convergence and the numerical results illustrate its performance in relation to existing approaches, without increasing complexity. 
The simulations showed that the performance of the proposed AR-SBP decoder for polar codes is better than that of BP and SC decoders, and is close to that of the more costly SCL decoders.

\end{document}